\documentstyle[11pt]{article}

\newcommand{\bigzerou}{%
\smash{\lower1.7ex\hbox{\bg 0}}}
\newcommand{\beq}{\begin{equation}}
\newcommand{\enq}{\end{equation}}

\newcommand{\mapright}[1]{%
\smash{\mathop{%
\hbox to 1.0cm{\rightarrowfill}}\limits^{#1}}}
\newcommand{\mapleft}[1]{%
\smash{\mathop{%
\hbox to 1.3cm{\leftarrowfill}}\limits^{#1}}}

\newcommand{\no}{\nonumber}

\newcommand{\beqy}{\begin{eqnarray}}
\newcommand{\enqy}{\end{eqnarray}}

\begin{document}


\begin{titlepage}
\vglue 3cm

\begin{center}
\vglue 0.5cm
{\Large\bf Gap Condition and Self-Dualized ${\cal N}=4$ Super Yang-Mills Theory for $ADE$ Gauge Group 
on $K3$}
\vglue 1cm
{\large Toru Sasaki} 
\vglue 0.5cm
{\it Department of Physics,
Hokkaido University, Sapporo 060-0810, Japan}\\
and \\
{\it Graduate School of Mathematics,
Nagoya University, Nagoya, 464-8602, Japan}\\ 
{ sasaki@particle.sci.hokudai.ac.jp}

\baselineskip=12pt

\vglue 1cm
\begin{abstract}
 We try to determine the partition function of 
${\cal N}=4$ super Yang-Mills theory for $ADE$ gauge  group on $K3$
by self-dualizing our previous $ADE$ partition function.
The resulting partition function satisfies gap condition.
\end{abstract}
\end{center}
\end{titlepage}
\section{Introduction}
\label{sec:1}
\setcounter{equation}{0}

SUSY duality has been giving a deep understanding
of non-perturbative phenomena
not only in physics but also in mathematics,
with fundamental connection to string duality.
${\cal N}=4$ super Yang-Mills theory is most suitable laboratory
to study this duality,
since ${\cal N}=4 $ is maximal SUSY in 4-dimentional theory without gravity,
and conformal invariant.
Indeed, many concrete and precise
calculations of ${\cal N}=4$ SUSY models were carried out 
and contributed to the deep understanding of duality. 

Vafa and Witten first attempted to test duality conjecture
of ${\cal N}=4$ super Yang-Mills theory on 4-manifold by determining the exact partition function\cite{vafa-witten}.
To determine the exact partition function,
some additional condition must be considered.
Firstly we need to use
the twisted ${\cal N}=4$ super Yang-Mills theory
instead of ${\cal N}=4$ super Yang-Mills theory itself.
To include the contribution from the strongly coupled region,
it is convenient to consider topologically twisted theory \cite{vafa-witten,laba}.
Secondly we need to consider this twisted ${\cal N}=4$ super Yang-Mills theory
on  K\"ahler 4-manifold
so that this twisted ${\cal N}=4$ theory can be reduced to ${\cal N}=1 $
theory by adding mass term proportional to the section of the canonical bundle.
As far as we are only interested in observables (such as correlation functions
or partition functions), it reproduces the same results completely. 
${\cal N}=1 $ theory is 
easier than 
${\cal N}=4 $ or ${\cal N}=2 $ theory
to determine the partition function.
Thus we consider the twisted ${\cal N}=4$ super Yang-Mills theory on
K\"ahler 4-manifold such as $C{\bf P}^2,{\hat C{\bf P}}^2,ALE,T^4,K3$ and $\frac{1}{2}K3$. 
It is well-known that duality conjecture
of ${\cal N}=4$ super Yang-Mills theory on 4-manifold is believed to be Montonen-Olive duality \cite{M-O}. Montonen-Olive duality is a kind of strong/weak duality($S$-duality). 
In addition to this duality, Montonen-Olive duality accompanies 
an exchange of gauge group (${\cal G}\leftrightarrow \hat{\cal G}$ ).
Here $\hat{\cal G}$ is the dual gauge group of ${\cal G}$.

As for $SU(N)$ theory on $K3$, Vafa-Witten derived
the exact partition, which satisfies the sharpened version of 
Montonen-Olive conjecture \cite{vafa-witten, lozano, m-v}.
On the other hand, there has been almost no attempt to determine
the partition function of $D,E$ gauge theory 
in spite of the fact that $S$-duality conjecture for these cases
has already been established \cite{vafa,M-O}. 
Thus we tried to derive the $D,E$ partition function on $K3$
in the previous article \cite{jin3}.
Fundamental strategy is the following:
We speculated a piece of $D,E$ partition function on $K3$,
which was derived by using the denominator identity of affine Lie algebra
for $ADE$ group \cite{kac, mac}.
Next we produced a set of functions derived by
modular transformation of this piece.
Finally we constructed the $D,E$ partition function on $K3$
by combining these functions,
so that these functions 
reproduced Montonen-Olive duality.
The resulting partition function satisfied Montonen-Olive duality 
completely !

Our $D,E$ partition function has serious problem.
This partition function does not satisfy the gap condition \cite{vafa-witten}.
Roughly speaking the gap 
comes from the irreducibility of the connections \cite{AHS},
and 
is necessary for ${\cal N}=4$ partition functions \cite{vafa-witten}.
There is similar situation in $U(N)$ partition functions on $\frac{1}{2}K3$ \cite{m-n, m-v, E-S}.
In this case this problem was solved by holomorphic anomaly.  
In the technical sense, the success of
the solution of this problem
is based on the self-duality of $U(N)$ theory.
Then $U(N)$ partition function is described by quasi-modular forms.
In this case one only adds quasi-modular forms,
so that the partition function satisfies the gap condition \cite{m-n}.
On the other hand, since $U(N)=U(1)\times SU(N)/{\bf Z}_N$,
$U(N)$ partition function is decomposed
into U(1) flux part and $SU(N)/{\bf Z}_N$ part \cite{m-v}.
In this article,
we construct 
holomorphic part of $U(N)$ partition function on $K3$.
Strictly speaking, we construct self-dual $U(N)$ partition function,
by multiplying  $SU(N)/{\bf Z}_N$ partition function given by Vafa-Witten 
\cite{vafa-witten, m-v, lozano}
and $U(1)$ flux part together.
The resulting partition function is also described by modular forms.
Here we explain that it is possible 
to construct the self-dualized $D,E$ partition function on $K3$ 
from our previous results.
$SU(N)/{\bf Z}_N$ partition functions are classified by 't Hooft flux $v\in H^2(K3,{\bf Z}_N)$. Furthermore on $K3$, these partition functions are classified by
$v^2$,
which is called an orbit. 
Then $SU(N)/{\bf Z}_N$ partition function on $K3$ have an orbit structure.
The orbit structure is equivalent to Montonen-Olive duality.
Our $D,E$ partition functions on $K3$ also have this orbit structure.
Thus we construct the self-dualized $D,E$ partition functions on $K3$,
by multiplying the $D,E$ partition functions
by $U(1)$ flux part.
The resulting self-dualized $D,E$ partition functions are
also described by modular forms.
What kinds of modular forms can we add to this partition function ?
Again we get back to the $U(N)$ theory on $\frac{1}{2}K3$ \cite{m-n}.
In analyzing the holomorphic part of $U(N)$ partition function
on $\frac{1}{2}K3$ in $N$: odd prime case,
we observe that the holomorphic part of 
$U(N)$ partition function can be expressed by Hecke transformation of
order $N$ of $U(1)$ partition function
plus (polynomial of $j$ function $)\times U(1)$ partition function.
This form is very suggestive.
We apply this decomposition to $D,E$ on $K3$ case
and construct gapful self-dualized $D,E$ partition 
functions,
by adding (polynomial of $j$ function $)\times U(1)$ partition function
on $K3$.
At last we find out that the resulting gapful $D,E$ partition functions
have the same form as $U(N)$ partition functions with the same top $q$ term.
That is, we lose the characteristics of Montonen-Olive duality of each groups.
However we expect that we can reconstruct  
$D,E$ partition functions with each Montonen-Olive duality,
if we 
separate $U(1)$ flux part from total  gapful $D,E$ partition functions.
Separation of $U(1)$ flux part is remaining problem.

The organization of this article is the following:
In Sec.2 we review the Vafa-Witten theory 
mainly concentrated on the $SU(N)$ theory.
Important gap condition is also introduced. 
In Sec.3 we briefly sketch our previous results.
In Sec.4 we introduce the concept of the self-dualization,
by using 
the orbit structure of ${\cal N}=4~SU(N)$ theory on $K3$.
We also discuss the ${\cal N}=4~SU(N)$ theory on $\frac{1}{2}K3$
in this context.
In Sec.5 we construct the self-dualized $D,E$ partition function on $K3$.
Then we solve the gap condition for these 
self-dualized $D,E$ partition function on $K3$.
In Sec.6 we conclude and discuss the remaining problems.

\section{Review of Vafa-Witten Theory}
\label{sec:2}
\setcounter{equation}{0}
We believe that $S$-duality of the twisted ${\cal N}=4$ super Yang-Mills theory 
is Montonen-Olive duality like ${\cal N}=4$ super Yang-Mills theory itself.
According to Montonen-Olive duality \cite{M-O}, ${\cal G}$ theory is dual to ${\hat {\cal G}} $
theory. As for $ADE $ gauge theory, ${\hat {\cal G}}={\cal G}/\Gamma_{\cal G}$.
Here $\Gamma_{\cal G} $ is the center of this group and given by the following table.
\begin{table}[h]
\begin{center}
\begin{tabular}{|c|c|c|}
\hline
${\cal G}$& $\Gamma_{\cal G}$\\
\hline
$A_{N-1}$ & ${\bf Z}_N$ \\
\hline
$D_{2N}$ &  ${\bf Z}_2\times{\bf Z}_2$\\
\hline
$D_{2N+1}$ &  ${\bf Z}_4$\\
\hline
$E_r, r=6,7,8$ &  ${\bf Z}_{9-r}$\\
\hline
\end{tabular}
\end{center}
\end{table} \\
For the theory with dual gauge group ${\cal G}/\Gamma_{\cal G}$,
one can introduce 't Hooft flux $v\in H^2(X,\Gamma_{\cal G})$ to  classify ${\cal G}/\Gamma_{\cal G}$ theory.

\subsection{Vafa-Witten conjecture and Gap Condition}
In this subsection we denote celebrated Vafa-Witten conjecture
concerned with partition function of twisted ${\cal N}=4$ super Yang-Mills theory
on 4-manifold $X$ \cite{vafa-witten, lozano, yoshi}.
Vafa and Witten pointed out that
the partition function of twisted ${\cal N}=4$ super Yang-Mills theory
can have remarkable simple form,
if $X$ is a K\"ahler 4-manifold and vanishing theorem holds \cite{vafa-witten}.  
In this situation, its partition function has the form of 
the summation of the Euler number of the moduli space of the ASD equations.
More precisely,  for twisted ${\cal N}=4~ {\cal G}/\Gamma_{\cal G}$ theory
with 't Hooft flux $v\in H^2(X,\Gamma_{\cal G})$ on $X$,
the partition function of this theory is given by the formula 
\begin{equation}
Z^X_v(\tau):= q^{-{\frac{(r+1)\chi(X)}{24}}}\sum_k \chi({\cal M}(v,k))q^k
\;\;\;(q:=\exp(2\pi i \tau)),\label{zxvt}
\end{equation}
where ${\cal M}(v,k) $ is the moduli space of ASD connections 
associated to ${\cal G}/\Gamma_{\cal G}$-principal bundle with 't Hooft flux $v$ and fractional instanton number $k\in \frac{1}{2|\Gamma_{\cal G}|}{\bf Z}$.
In (\ref{zxvt}), $\tau$ is the gauge coupling constant including theta angle,
$\chi(X)$ is Euler number of $X$ and $r$ is the rank of ${\cal G} $.
$q^{-\frac{1}{24}} $ factor is required by modular property like 
the case of $\eta(\tau) $ function.
For later use, we introduce 
the following form
corresponding to $ {\cal G}$ 
theory itself
for $v=0$: 
$Z_t^X(\tau):=Z_{v=0}^X(\tau)=|\Gamma_{\cal G}|Z_{\cal G}^X(\tau) $.
$|\Gamma_{\cal G}|$ is the number of the elements of $\Gamma_{\cal G}$.
Furthermore we also introduce 
the following form
corresponding to 
$ {\cal G}/\Gamma_{\cal G}$ for its dual pair :
$Z_{{\cal G}/\Gamma_{\cal G}}^X(\tau):=\sum_vZ_v^X(\tau)$.

With this result, Vafa and Witten
conjectured the behavior of the partition functions under the action of $SL(2,{\bf Z})$ on $\tau$ .They started with 't Hooft's work \cite{tHooft} in mind. 
In \cite{tHooft}, the path integral with ${\bf Z}_N$-valued electric flux and 
that with magnetic flux 
are related by Fourier transform.
Vafa and Witten combined the conjecture of strong/weak duality to this
't Hooft's result. Their conjecture is summarized by the following formula:  

\begin{equation}
Z^X_t\left(-\frac{1}{\tau}\right)=\vert\Gamma_{\cal G}\vert^{-\frac{b_2(X)}{2}}
\left(
\frac{\tau}{i}
\right)^{-\frac{\chi(X)}{2}}
Z_{{\cal G}/\Gamma_{\cal G}}^X(\tau).
\label{m-o}
\end{equation}

In $v=0$ case
${\cal M}(0,k):={\cal M}_k^{\cal G}  $ is  the moduli space of irreducible 
ASD connections associated to ${\cal G}$-principal bundle with instanton number $k$.
Its dimension $\dim {\cal M}(0,k) $ is given by
Atiyah-Hitchin-Singer dimension formula \cite{AHS}:
 
\beq
\dim{\cal M}^{\cal G}_k=4h({\cal G})(k-r)-4r.
\enq 
where $h({\cal G})$ is the dual Coxeter number, $\dim{\cal G}$ is the dimension
of ${\cal G}$ and $k$ is the instanton number. 
$ \dim{\cal M}^{\cal G}_k\ge 0$ means $k\ge r+1$.
That is, the moduli space of irreducible ASD connections with 
$ADE$ gauge group on $K3$ can exist only for $k\ge r+1$ except for $k=0$ trivial case.
This condition restricts the form of the partition function. 
This condition is called gap condition \cite{vafa-witten, m-n, m-v, E-S, jin3}.

\subsection{$SU(N)$ partition function on $K3$}
In this section we review Vafa-Witten's results for $SU(N)$ theory on $K3$
 \cite{vafa-witten, m-v, lozano}.
As mentioned above, $K3$ is a good manifold
to determine the partition function.
Especially there is a brane picture of ${\cal N}=4~ U(N)$ gauge theory on $K3$ \cite{m-v, lozano}.  
By using the picture that ${\cal N}=4 ~U(1)$ gauge theory is given 
by single $M5$ brane wrapped around $K3\times T^2$,
one can understand that $U(1)$ partition function on $K3$
is given by \cite{vafa-witten, m-v, lozano}: 

\beq
\frac{1}{\eta^{24}(\tau)}:=G(\tau).
\enq
Here we ignore the contribution from the $U(1)$ flux.

To obtain ${\cal N}=4 ~U(N)$ gauge theory on $K3$,
we consider $N$ coincident $M5$ brane wrapped around $K3\times T^2$.
Furthermore by fixing 't Hooft flux with trivial $v=0$,
we obtain  ${\cal N}=4 ~SU(N)$ gauge theory on $K3$.
In this picture, Hecke transformation naturally appears
as the effect of summing up all the ways of wrapping $T^2$
in $K3\times T^2$ by $T^2$ in $M5$ brane.
Based on this idea we define the following $SU(N)$ partition function: 
\beqy
Z_{SU(N)}(\tau)&:=&\frac{1}{N}Z_t(\tau)
\no\\
&=&
\frac{1}{N^3}\sum_{0\le a,b,d \in{\bf Z} \atop ad=N,b<d}
dG(\frac{a\tau+b}{d}).
\label{zpmt}
\enqy
This is almost Hecke transformation of order $N$ of $G(\tau)$.
Total $U(N)$ partition function based on this picture are derived in
\cite{bonelli, m-v}.
But the formula (\ref{zpmt}) is not true Hecke transformation. 
This subtle difference is the origin of Montonen-Olive duality
for $SU(N)$ as will be explained later. Next
its dual pair $SU(N)/{\bf Z}_N$ partition function on $K3$
is simply obtained by modular transformation $\tau\to -\frac{1}{\tau}$
and combined with Montonen-Olive duality (\ref{m-o}).
Throughout this section we avoid to mention the notion of orbit
which corresponds to Hecke transformation.
In forth section we show this notion in detail.
Furthermore we explain the trick to produce
Montonen-Olive duality from (\ref{zpmt}).
Finally we note that the partition function
trivially satisfies the gap condition.
\section{Our Previous Results}
\label{sec:2}
\setcounter{equation}{0}
$K3$ has many special properties.
One is its orbifold construction.
$K3$ can be constructed by the following processes.
First we divide 4-torus $T^4$ by ${\bf Z}_2$ and obtain 
quotient space $T^4/{\bf Z}_2:=S_0$.
Next we blow up its sixteen singularities by ${\cal O}(-2) $ curve
and obtain smooth K\"ahker surface $K3$.
$K3$ has the following data \cite{Fukaya}:
\beq
\chi(K3)=24, K_{K3}={\cal O},\mbox{Intersection form of~} K3~ \mbox{is~}(-E_8)^{\oplus 2}\oplus H^{\oplus 3}.
\enq
Here $K_{K3}$ is canonical bundle of $K3$ and ${\cal O}$ stands for trivial.
\[
H=\left(\begin{array}{cc}
0&1\\
1&0
\end{array}\right).
\]
We tried to reproduce the above geometrical processes in the partition function
level for $SU(N)$ \cite{jin, jin2}. The trial was successful, that is, $SU(N)$ partition on $K3$
has the product of two factors.
The first one is the contribution from $S_0$ and has the form:
Hecke transformation of order $N$ of $1/\eta^8(\tau)$.
The second one is the contribution 
obtained from blowing up sixteen singularities
and the form
is given by
the blow-up formula \cite{yoshioka, kap, vafa-witten}. 
These two factors are summed up so that the resulting partition function
satisfies Montonen-Olive duality.
Our partition function is consistent to the fact that
$SU(N)$ partition function on $K3$ is the Hecke transformation of order $N$
of $1/\eta^{24}(\tau)$ \cite{vafa-witten, m-v, yoshihecke, lozano, mukai, nak}.
Key point of this verification is  that eta function $\eta(\tau/N) $
can be interpreted as blow up formula   
which has typically the form $\theta(\tau)/\eta^m(\tau)$ \cite{yoshioka, kap}.
Fortunately there are beautiful identities \cite{jin2}:
\beq
\frac{1}{\eta(\frac{\tau}{N})}=\frac{\theta_{A_{N-1}}(\tau)}{\eta^N(\tau)}.
\label{moa}
\enq
These identities are already verified by celebrated denominator identity \cite{kac, mac}.

In the next work, we tried to determine the ${\cal N}=4~ ADE$ partition function
on $K3$ \cite{jin3}.
Using the generalization of (\ref{moa}) to $ADE$, 
we define the $ADE $ blow-up formula:
\beq
 \frac{\theta_{{\cal G}_r}(\tau)}{\eta^{r+1}(\tau)}=\mbox{eta product},
 \label{di}
\enq
where we note that these blow-up formula also have the form of eta product
and this fact is also verified by denominator identity \cite{kac, mac}.
By requiring the duality conjecture to these eta product,
we want to construct ${\cal N}=4 ~ADE$ partition on $K3$.
Before moving to concrete construction,
we explain why it is reliable to adopt $ADE$ blow-up formula
to construct ${\cal N}=4 ~ADE$ partition on $K3$.
There is a stringy picture:

\begin{tabular}{ccc}
\\
IIA on $K3\times T^2 \times ALE_{\small ADE}$ &$\leftrightarrow$& Hetero on $T^4\times T^2 \times ALE_{ADE}$\\
$\downarrow$ & &$\downarrow$\\
${\cal N}=4 ~ADE$ on $K3$ &$\leftrightarrow$& (${\cal N}=4~ U(1)$ on $ALE_{ADE})^{\otimes 24}$
\\
\\
\end{tabular}\\
First line is IIA/Hetero duality \cite{hj}. To second line in IIA side, we compactify
$T^2\times ALE_{ADE}$ and obtain ${\cal N}=4 ~ADE$ on $K3$.
To the second line in Hetero side, we compactify $T^4\times T^2$
and obtain $({\cal N}=4~U(1)$ on $ALE_{ADE})^{\otimes 24}$ \cite{jin3}.
The origin of ${\otimes 24}$ is explained by 
the compactification of $K3\times T^2$
in IIA side.
Thus each side of 
the second line is equivalent and their partition functions 
in both side are the same.
On the other hand, the partition function of ${\cal N}=4~U(1)$ on $ALE_{ADE}$
was already obtained by Nakajima \cite{naka}.
His results are very similar to ours in (\ref{di}).
Thus we adopt  24-th power of (\ref{di}) as a piece of 
${\cal N}=4~ADE$ partition function on $K3$.
To construct  ${\cal N}=4~ADE$ partition function on $K3$ from (\ref{di}),
we introduce 24-th power of (\ref{di}),which we call primary function.
First we generate a set of functions,
by modular transformation of the primary function
(and adding to $\tau \to \tau+1/m $ transformation in some cases).
Next we request Montonen-Olive duality to appropriate
liner combination of these functions
and determine the coefficient of these functions.
In practice, it is difficult to produce the partition functions
directly from these functions.
We use the following observation.
We think of 
a subset of the functions.
Appropriate liner combination of functions of this subset 
has the same modular property 
that a
piece of $SU(N)$
partition function has \cite{jin3}.
Here ${\cal G}_r$ and $SU(N)$ have the same Montonen-Olive duality.
We call a piece of $SU(N)$
partition functions as $G_j(\tau)$
and appropriate liner combination of functions of this subset as ${\tilde G}_j(\tau)$.
Then modular property of $G_j(\tau)$ and ${\tilde G}_j(\tau)$
is completely the same.
For ${\tilde G}_j(\tau)$, we use the same coefficient
as that of $G_j(\tau)$ in $Z_t(\tau)$ or $Z_{SU(N)/{\bf Z}_N}(\tau)$.
Finally we obtain the ${\cal N}=4~ADE$ partition function
satisfying Montonen-Olive duality.
Unfortunately the resulting partition functions does not satisfy
the gap condition.
We have to solve this problem.

\section{$U(N)$ Theory on $K3$ and $\frac{1}{2}K3$}
\label{sec:3}
\setcounter{equation}{0}

\subsection{Orbit, Hecke and Montonen-Olive duality}
In this part, we will explain the orbit structure of $SU(N)$,
by using examples of $SU(2)$ or $SU(4)$.

On $K3$, $ {\cal G}/\Gamma_{\cal G}$ partition function with
't Hooft flux $v\in H^2(K3,\Gamma_{\cal G})$
is classified by $v^2$ mod $2|\Gamma_{\cal G}|$ or $v=0$
instead of $v$ itself in $|\Gamma_{\cal G}|$:prime case 
\cite{vafa-witten,jin3},
because instanton number $k$ is given by $k=n-v^2(N-1)/2N$ 
in $SU(N)/{\bf Z}_N$ case.
We call $v^2$ as an orbit.
In $\Gamma_{\cal G}={\bf Z}_2 ({\cal G}=SU(2)$ or $E_7$ case) for instance,
$2^{22} Z_v(\tau)$'s are classified into three types
$v^2=2j$ mod $4 (j=0,1)$ or $v=0$.
The numbers of these types of orbits are counted as \cite{vafa-witten}
  \begin{eqnarray}
n_0&=&
\mbox{number of }v^2\equiv 0 \mbox{ mod }4 \mbox{ and }v\ne 0
\no
\\ 
&=&
2^{21}+2^{10}-1,\\
n_1&=&
\mbox{number of }v^2\equiv 2j \mbox{ mod }4 
\no
\\ 
&=&
2^{21}-2^{10},\\
n_t&=&
\mbox{number of }v=0 
\no
\\ 
&=&1.
\end{eqnarray}
In $\Gamma_{\cal G}={\bf Z}_4 ({\cal G}=SU(4)$ or $D_{2N+1}$ case) for instance,
$4^{22} Z_v(\tau)$'s are first classified into five types
$v^2=2j$ mod $8(j=0,\ldots,3,)$ or $v=0$.
Next we think of $v=2{\tilde v}$ case.
Then $v$ are trivially classified into $v^2 =0  $ mod $8$,
but we furthermore think 2 types ${\tilde v}^2 =2j  $ mod $4(j=0,1,)$.
Totally there are 7 types of orbits, and the numbers of orbits are counted as \cite{jin3}
\begin{eqnarray}
n_0^{4}&=&2^{42}+2^{31}-2^{21},\no\\
n_1^{4}&=&2^{42}-2^{31},\no\\
n_2^{4}&=&2^{42}+2^{31}-2^{21},\no\\
n_3^{4}&=&2^{42}-2^{31},\no\\
n_0^{2}&=&2^{21}+2^{10}-1,\no\\
n_1^{2}&=&2^{21}-2^{10}\no,\\
n_t^1&=&1.
\end{eqnarray}

We will show how $n_j$ work in the determination of the partition function.
We pick up $SU(2)$ case as an example.
These numbers $n_j$'s  come into duality conjecture
\beq
Z_t(-\frac{1}{\tau})=\tau^{-12}2^{-11}Z_{SU(2)/{\bf Z}_2}(\tau),
\label{zsu2}
\enq 
in 
\beq
Z_{SU(2)/{\bf Z}_2}(\tau)=\sum_v Z_v(\tau)=n_t Z_t(\tau)+n_0Z_0(\tau)+n_1Z_1(\tau).
\enq
If we think of the form
\beq
Z_t(\tau)=a\frac{1}{\eta^{24}(2\tau)}+\frac{1}{2}\left(
\frac{1}{\eta^{24}(\frac{\tau}{2})}+\frac{1}{\eta^{24}(\frac{\tau+1}{2})}\right),
\enq
\beq
Z_j(\tau)=\frac{1}{2}\left(
\frac{1}{\eta^{24}(\frac{\tau}{2})}+\frac{(-1)^j}{\eta^{24}(\frac{\tau+1}{2})}\right),
\enq
and request the modular property (\ref{zsu2}), then
we have $a=\frac{1}{4}$ and 
can determine the partition function
\beq
Z_t(\tau)=\frac{1}{4}\frac{1}{\eta^{24}(2\tau)}+\frac{1}{2}\left(
\frac{1}{\eta^{24}(\frac{\tau}{2})}+\frac{1}{\eta^{24}(\frac{\tau+1}{2})}\right).
\label{zt}
\enq
This is the formula as we have already seen in (\ref{zpmt}).
But this is not true Hecke transformation of order $2$ of $1/\eta^{24}(\tau)$ with modular weight $-12$.
Here we introduce true Hecke transformation of order $N$ of $\phi^k(\tau)$ with modular weight $k$ \cite{m-v}:
\beq
T_N \phi^k(\tau):= N^{k-1}\sum_{0\le a,b,d \in{\bf Z} \atop ad=N,b<d}
d^{-k}\phi^k(\frac{a\tau+b}{d}).
\enq
Using this, we can introduce,
\beq
2^{11}T_2 \frac{1}{\eta^{24}(\tau)}=\frac{1}{4}\frac{1}{\eta^{24}(2\tau)}+\frac{2^{11}}{2}\left(
\frac{1}{\eta^{24}(\frac{\tau}{2})}+\frac{1}{\eta^{24}(\frac{\tau+1}{2})}\right).\label{t2}
\enq
(\ref{t2}) has $2^{11}$ times degeneracy of $Z_t(\tau)$
and
$
Z_0(\tau)$.
$2^{11}$ has the same origin of $2^{-11}$ in (\ref{zsu2}). 
If we change true Hecke (\ref{t2}) into (\ref{zt}),
then the degeneracy is resolved and 
the partition function of 
each orbit acquires the property of Montonen-Olive duality of $SU(2)$ theory on $K3$ \cite{vafa-witten}.

\subsection{$U(N)$ on $K3$}

Hereafter we consider
the partition function of self-dualized $A_{N-1}$ theory on $K3$.
This is nothing but $U(N)$ partition function on $K3$.
For this purpose, we multiply $SU(N)/{\bf Z}_N$ partition function of each orbit
by the corresponding $U(1)$ flux part and obtain $U(N)$ partition function. 
In $U(1)$ case, the $U(1)$ flux part is given by \cite{m-v}:
\beq
\theta_{\Gamma^{19,3}}(\tau,{\bar \tau})=\sum_{P_L,P_R\in \Gamma^{19,3}}
q^{\frac{1}{2}P_L^2}{\bar q}^{\frac{1}{2}P_R^2},
\label{t193}
\enq
where $\Gamma^{19,3} $ is 
the even, self-dual, integral lattice with signature $(19,3)$ (Narain lattice)
and $P_L,P_R$ are polarization on it.
The total partition function with the $U(1)$ flux part
is given by \cite{m-v}:
\beq
Z_1(\tau,{\bar \tau})=\theta_{\Gamma^{19,3}}(\tau,{\bar \tau})G(\tau).
\label{z1}
\enq 
Modular weight of this partition function is $(19/2,3/2)+(-12,0)=(-5/2,3/2)$.
For the following reason, we want to remove ${\bar \tau}$-dependence 
from this partition function. In the next section, we discuss the gap condition
of self-dualized $D,E$ partition function with $U(1)$ flux part.
In this case if we multiply 
our previous $D,E$ partition function by
${\bar \tau}$-dependent $U(1)$ flux part
and intend to fulfill the gap condition,
then we must consider the possibility for these partition functions to include 
$E_2$-depnedent terms.
But $E_2$-depnedent terms request large ambiguity 
to cancel the negative $q$-powers except for top $q$ term.
So in technical reason,
we want to restrict the holomorphic functions
to cancel the negative $q$-powers.
Thus before discussing the gap condition,
we consider the $U(1)$ flux part 
by keeping the partition function holomorphic.
We remove ${\bar \tau}$-dependence from (\ref{t193}).
The following derivation is the generalization of $U(1)$ on $\frac{1}{2}K3$ in \cite{m-v}.
Remembering $\Gamma^{19,3}=\Gamma^{16}\oplus \Gamma^{3,3}$ and $(P_L,P_R)=(P_L^\prime,0)\oplus({\tilde P}_L,{\tilde P}_R)$,
we separate ${\bar \tau}$-dependence from (\ref{t193}),
\beq
\theta_{\Gamma^{19,3}}(\tau,{\bar \tau})=\sum_{{\tilde P}_L,{\tilde P}_R\in \Gamma^{3,3}}q^{\frac{1}{2}{\tilde P}_L^2}{\bar q}^{\frac{1}{2}{\tilde P}_R^2}\cdot \sum_{P_L^\prime\in \Gamma^{16}}q^{\frac{1}{2}P_L^{\prime 2}},
\enq
where we note that $\Gamma^{3,3}=(\Gamma^{1,1})^{\oplus 3}$ 
and $\Gamma^{16}=(\Gamma^{8})^{\oplus 2}$ and introduce
\begin{eqnarray}
\sum_{{\tilde P}_L,{\tilde P}_R\in \Gamma^{3,3}}q^{\frac{1}{2}{\tilde P}_L^2}{\bar q}^{\frac{1}{2}{\tilde P}_R^2}&:=&\theta_H^3(\tau,{\bar \tau}),
\\
\sum_{P_L^\prime\in \Gamma^{16}}q^{\frac{1}{2}P_L^{\prime 2}}&:=&\theta_{\Gamma^8}^2(\tau)=E_4^2(\tau).
\end{eqnarray}
Here in general 
$E_k(\tau)$ is $k$-th Eisenstein series with modular weight $k$.
Remembering $K3$ is elliptic surface \cite{Fukaya}, we put the size of this elliptic fiber $1/R$.
Then choosing a metric, we can put
\beq
({\tilde P}^j_L,{\tilde P}^j_R)=(\frac{m^j+n^j}{2R}+\frac{(m^j-n^j)R}{2},\frac{m^j+n^j}{2R}-\frac{(m^j-n^j)R}{2})~,j=1,2,3.
\enq
$\theta_H(\tau,{\bar \tau}) $ is given by
\beq
\theta_H(\tau,{\bar \tau})=\sum_{m,n\in {\bf Z}}q^{\frac{1}{2}(\frac{m+n}{2R}+\frac{(m-n)R}{2})^2}{\bar q}^{\frac{1}{2}(\frac{m+n}{2R}-\frac{(m-n)R}{2})^2}.
\label{tH}
\enq
We take the limit $1/R\to 0 $,where the size of the elliptic fiber shrinks to zero. Then $m=n$ only contributes to (\ref{tH}) and the summation of  (\ref{tH})
can be carried out as:
\beq
\theta_H(\tau,{\bar \tau})=\frac{R}{\sqrt{2 Im \tau}}.
\enq 
Note that modular weight of $1/\sqrt{Im \tau} $ is $(1/2,1/2)$.
In the limit $1/R\to 0 $,we can extract holomorphic part $\theta_{\Gamma^{8}}^2(\tau)=E_4^2(\tau) $ from $\theta_ {\Gamma^{19,3}}(\tau,{\bar \tau})$.
We redefine the total $U(1)$ partition function by
\beq
{\hat Z}_1(\tau):=\theta_{\Gamma^8}^2(\tau)G(\tau)=\frac{E_4^2(\tau)}{\eta^{24}(\tau)}.
\enq

We proceed to $U(2)$ theory. In this case we multiply $SU(2)/{\bf Z}_2$ partition function with $v$ by the corresponding $U(1)$ flux part,
where we introduce $v=(V,(i^k,j^k)),k=1,2,3$ corresponding to $\Gamma^{19,3}=\Gamma^{16}\oplus \Gamma^{3,3}$.
In the same way as $U(1)$ case, if we take the limit $1/R\to 0$, then
only $i^k=j^k$ contribute to the partition function and the partition function
only depends on $V$. This yields $2^3$ times degeneracy 
to be compared with 
original $SU(2)/{\bf Z}_2$ partition function. Thus we change $SU(2)$ partition function
\beq
Z_t(\tau)=\frac{1}{4}G(2\tau)+\frac{1}{2}\left(G(\frac{\tau}{2})+G(\frac{\tau+1}{2})\right)
\enq
into
\beq
{\hat Z}_t(\tau)=\frac{1}{4}G(2\tau)+\frac{2^3}{2}\left(G(\frac{\tau}{2})+G(\frac{\tau+1}{2})\right).
\label{zh2}
\enq
$Z_j(\tau) \to {\hat Z}_j(\tau)=2^3 Z_j(\tau)$ is also done.
Since $2^{11}=2^3\cdot 2^{8}=2^3(1+2^8-1)$, 
$2^8$ times degeneracy of
$2^{11}T_2 G(\tau)$ in (\ref{t2}) 
is resolved into (\ref{zh2}). Indeed corresponding Montonen-Olive duality is
\beq
{\hat Z}_t(-\frac{1}{\tau})=\tau^{-12}2^{-8}{\hat Z}_{SU(2)/{\bf Z}_2}(\tau).
\enq
What about $U(1)$ flux part of $U(2)$ ? These are given by
\begin{eqnarray}
{\hat \theta}_t(\tau)&:=&4E_4^2(2\tau),\\
{\hat \theta}_0(\tau)&:=&2\left(E_4^2(\frac{\tau}{2})+E_4^2(\frac{\tau+1}{2})\right)-4E_4^2(2\tau),\\
{\hat \theta}_1(\tau)&:=&2\left(E_4^2(\frac{\tau}{2})+E_4^2(\frac{\tau+1}{2})\right).
\end{eqnarray}
Refer to $U(1)$ flux part on $\frac{1}{2}K3$ in \cite{m-v}.
To obtain $U(2)$ partition function, we multiply $U(1)$ flux part by the corresponding $SU(2)/{\bf Z}_2$ partition function, so that $U(2)$ partition function is self-dual. We finally obtain
\begin{eqnarray}
{\hat Z}_2(\tau)&:=&{\hat \theta}_t(\tau){\hat Z}_t(\tau)+{\hat \theta}_0(\tau){\hat Z}_0(\tau)+{\hat \theta}_1(\tau){\hat Z}_1(\tau)
\\
&=&E_4^2(2\tau)G(2\tau)+2^4\left(E_4^2(\frac{\tau}{2})G(\frac{\tau}{2})+E_4^2(\frac{\tau+1}{2})G(\frac{\tau+1}{2})\right)\\
&=&2^3T_2 {\hat Z}_1(\tau).
\end{eqnarray}
Indeed ${\hat Z}_2(-\frac{1}{\tau})=\tau^{-4}{\hat Z}_2(\tau)$.
The final expression is natural, since $U(2)$ partition function is 
the true Hecke transformation. 
The Hecke transformation changes modular form into modular form.  
We also note that this partition function trivially 
satisfies the gap condition.
This situation is true for any $N$ case.

Generalization to $U(N)$ is straightforward not only in $N$ ;prime case,
 but also in $N$; non-prime case. That is
\beq
{\hat Z}_N(\tau)=N^{3}T_N {\hat Z}_1(\tau).
\enq
For later use, we give explicit $U(4)$ partition function:
\begin{eqnarray}
{\hat Z}_4(\tau)&=&
E_4^2(4\tau)G(2\tau)+2^4\left(E_4^2(\frac{2\tau}{2})G(\frac{2\tau}{2})+E_4^2(\frac{2\tau+1}{2})G(\frac{2\tau+1}{2})\right)\no\\
&&+4^4\left(E_4^2(\frac{\tau}{4})G(\frac{\tau}{4})+E_4^2(\frac{\tau+1}{4})G(\frac{\tau+1}{4})+\cdots+E_4^2(\frac{\tau+3}{4})G(\frac{\tau+3}{4})\right)
\\
&=&4^3T_4 {\hat Z}_1(\tau),
\end{eqnarray}
\beq
{\hat Z}_4(-\frac{1}{\tau})=\tau^{-4}{\hat Z}_4(\tau).
\enq
In this subsection we derived holomorphic part of $U(N)$ partition function on $K3$,
by multiplying $SU(N)$ partition function and holomorphic part of $U(1)$ flux part together.
The final results are described by the Hecke operator.
While the standard $U(N)$ partition function on $K3$ includes
${\bar \tau}$-dependence, this partition function is also described by
the Hecke operator \cite{m-v,bonelli,yoshitrap}.
\subsection{$U(N)$ on $\frac{1}{2}K3$}

In this subsection, we investigate the holomorphic part of
$U(N)$ partition functions
on $\frac{1}{2}K3$
to derive the self-dualized $D,E$ partition function
satisfying the gap condition in the next section.
There are already results of $U(N)$ partition functions on $\frac{1}{3}K3$ \cite{m-v, m-n, E-S}.
In \cite{m-v} 
the authors thought the same kind of limit 
as in the previous subsection
and removed 
the contribution of $\Gamma^{1,1}$
from that of $\Gamma^{9,1}=\Gamma^8\oplus\Gamma^{1,1}$. 
Their partition function is formally 
a holomorphic
function. In fact their $U(1)$ partition function in \cite{m-v} is given by
\beq
G_1(\tau)=\frac{E_4(\tau)}{\eta^{12}(\tau)}.
\label{g1}
\enq
Here we note that (\ref{g1}) is quasi-modular instead of self-dual.
That is
\beq
G_1(\tau+1)=-G_1(\tau).
\enq 
This situation always occurs in $N$; odd case.
However from one method of their derivation of $U(N)$ partition functions;
the holomorphic anomaly equations and the gap condition,
the holomorphic anomaly or $E_2$-depnedence comes
in the $U(N)$ partition functions except for $N=1$ case.
Their explicit results for $G_3$ and $G_5$ in \cite{m-n} are given by 

\beq
G_3(\tau)=\frac{E_4}{576\eta^{36}}(54E_2^2E_4^2+109E_4^3+216E_2E_4E_6+197E_6^2),
\enq
\begin{eqnarray}
G_5(\tau)&=&\frac{E_4}{2985984\eta^{60}}(18750E_2^4E_4^4+136250E_2^2E_4^5+116769\no\\
&&
+150000E_2^3E_4^3E_6+653000E_2E_4^4E_6+426250E_2^2E_4^2E_6^2
\no\\
&&
772460E_4^3E_6^2+505000E_2E_4E_6^3+207505E_6^4).
\end{eqnarray}
These results of course have  $E_2$ terms.
But we are only interested in the holomorphic part or $E_2$-independent terms.
In our interpretation, we expect that 
the holomorphic part of
$G_N:N$ (odd prime) contains
Hecke transformation of $G_1$
and top $q$ term only comes from Hecke transformation.
This expectation comes from the fact that
 Hecke transformation changes quasi-modular forms $G_1(\tau)$
into quasi-modular form in $N$; odd prime case.
Then, what about residual of Hecke transformation of $G_1$ in holomorphic part of $G_N$ ?
We expect that these parts are expressed as
(polynomial of $j$ function) $\times G_1$
from the modular weight without no ambiguity.
Here $j(\tau)=E_4(\tau)^3/\eta(\tau)^{24}$.
Note that $j(\tau)$ and $G_1(\tau)$ have the only singularity at $i \infty$
on the upper half plane ($Im \tau >0$).
Finally we express the holomorphic part $G^{hol}_3,G^{hol}_5$ and $G^{hol}_7$
in \cite{m-n} as follows:
\beq
G_3^{hol}(\tau)=\frac{51}{32}T_3G_1(\tau)-\frac{495}{8}G_1(\tau),
\enq

\beq
G_5^{hol}(\tau)=\frac{913945}{497664}T_5G_1(\tau)-\frac{1994875}{41472}jG_1(\tau)+\frac{1617948125}{27648}G_1(\tau),
\enq

\begin{eqnarray}
G_7^{hol}(\tau)&=&\frac{19871110007}{10749542400}T_7G_1(\tau)-\frac{12533265619}{298598400}j^2G_1(\tau)
\no
\\
&&
+\frac{16274075386961}{199065600}jG_1(\tau)-\frac{37025797373687003}{1343692800}G_1(\tau).
\end{eqnarray}
The above expression are natural,
since we use Hecke transformation of $G_1$ 
and $j$ function instead of $E_4,E_6$
as independent modular forms.
But we want to use Hecke transformation of $G_1$,
which corresponds to the self-dualized $D,E$ partition function.

\section{Self-dualized $D,E$ Theory on $K3$}
\label{sec:4}
\setcounter{equation}{0}
To solve the problem of gap condition of ${\cal N}=4~D,E$
partition function on $K3$,
it is 
the best to treat $D,E$ partition function itself
and intend to cancel negative $q$ powers except for top $q$ term. 
But by now we have no guiding principle. 
So we self-dualize our $D,E$ partition function,
and concentrate on a gap condition
at the sacrifice of non-trivial Montonen-Olive duality.

\subsection{Self-Dualizing Our Original $D,E$ Partition Function on $K3$ }
For $D_{2N+1},E_r$, it is straightforward to self-dualize 
the original
$D,E$
partition function. As we have mentioned in Sec.3,
modular property of ${\tilde G}_j(\tau)$ and $G_j(\tau)$
is the same \cite{jin3}.
Thus we simply replace $G_j(\tau)$ by ${\tilde G}_j(\tau)$
and follow the process in Sec.4.
Then we obtain self-dualized $D,E$ partition function.
We write down general process for $D_{2N+1},E_8$ case.
\begin{eqnarray}
{\hat Z}_{UD_{2N+1}}(\tau)&:=&E^2_4(4\tau){\tilde G}^1_t(\tau)+2^{4}
\left(
E_4^2\left(\frac{2\tau}{2}\right){\tilde G}^2_0(\tau)
+E_4^2\left(\frac{2\tau+1}{2}\right){\tilde G}^2_1(\tau)
\right)
\no\\
&&
+4^{4}\left(+E_4^2\left(\frac{\tau}{4}\right){\tilde G}^4_0(\tau)
+E_4^2\left(\frac{\tau+1}{4}\right){\tilde G}^4_1(\tau)
+\cdots+E_4^2\left(\frac{\tau+3}{4}\right){\tilde G}^4_3(\tau)
\right),
\no\\
\end{eqnarray}
\beq
{\hat Z}_{UE_8}(\tau):=E_4^2(\tau)Z_{E_8}(\tau).
\enq
See explicit forms of
${\tilde G}_j(\tau) $ and $Z_{E_8}(\tau)$ in \cite{jin3}.

To self-dualize $D_{2N}$ partition function, we have to 
remind of
the discussion in \cite{jin3}.
In \cite{jin3}, the numbers of orbits of $D_{2N}$
is the same as those of $SU(2)$ on the manifold with $H^{\oplus 22}$.
To self-dualize $D_{2N}$ partition function
by $E_4^2(2\tau)$ etc,
we have to make degeneracy of original
$D_{2N}$ partition function 
by $2^{14}$ times.
Taking this point into consideration, we self-dualize $D_{2N}$ partition function as follows:
\beq
{\hat Z}_{UD_{2N}}(\tau):=E^2_4(2\tau){\tilde H}_0(\tau)+2^{16}
\left(
E_4^2\left(\frac{\tau}{2}\right){\tilde G}_0(\tau)
+E_4^2\left(\frac{\tau+1}{2}\right){\tilde G}_1(\tau)
\right).
\enq
$2^{16}$ factor reflects to the above point.

Here we list up several explicit self-dualized $D,E$ partition functions
by using the results in \cite{jin3}:

\begin{eqnarray}
{\hat Z}_{UD_2}(\tau)&=&\frac{E_4^2}{3888\eta^{72}}(323E_4^6+2594E_4^3E_6^2+971E_6^4)
\no\\
&=&
q^{-3}-24q^{-2}+780q^{-1}+116928+\cdots,
\end{eqnarray}

\begin{eqnarray}
{\hat Z}_{UD_4}(\tau)&=&\frac{E_4^2}{2902376448\eta^{120}}(24192611E_4^{12}
+638981788E_4^9E_6^2
\no\\
&&
+1548828426E_4^6E_6^4+653069452E_4^3E_6^6+7304171E_6^8)
\no\\
&=&q^{-5}+504q^{-3}-96q^{-2}+107370q^{-1}+116073216+\cdots,
\end{eqnarray}

\begin{eqnarray}
{\hat Z}_{UD_3}(\tau)={\hat Z}_4(\tau)&=&\frac{E_4^2}{5038848\eta^{96}}(135295E_4^{9}+2129577E_4^6E_6^2
\no\\
&&
+2473329E_4^6E_6^4+300647E_6^6)
\no\\
&=&q^{-4}+532728\cdots,
\end{eqnarray}

\begin{eqnarray}
{\hat Z}_{UD_5}(\tau)&=&\frac{E_4^2}{544195584\eta^{144}}(134710267E_4^{15}
+5375612131E_4^{12}E_6^2
\no\\
&&
+22295933590E_4^9E_6^4+20140185182E_4^6E_6^6+4165145615E_3^3E_6^8
\no\\
&&
+131189279E_6^{10})
\no\\
&=&48q^{-4}+25570944+\cdots,
\label{d5}
\end{eqnarray}

\begin{eqnarray}
{\hat Z}_{UD_7}(\tau)&=&\frac{E_4^2}{7487812485248974848\eta^{192}}(
3708321329732099E_4^{21}
+278349798105179209E_4^{18}E_6^2
\no\\
&&
+2495490821932031115E_4^{15}E_6^4+6027033285323884697E_4^{12}E_6^6
\no\\
&&
+4801092532525571089E_3^9E_6^8+1272963132360267843E_4^6E_6^{10}
\no\\
&&
+95766323147504225E_4^3E_6^{12}+1220755773779419E_6^{14})
\no\\
&=&2q^{-8}+2508q^{-4}+64q^{-2}+1371241872+\cdots,
\end{eqnarray}

\begin{eqnarray}
{\hat Z}_{UE_8}(\tau)&=&\frac{E_4^2}{1104122877824872835186688\eta^{216}}(
179571116071856545537E_4^{24}
\no\\
&&
+14834990989690717426744E_4^{21}E_6^2
+148775993053767172568668E_4^{18}E_6^4
\no\\
&&
+411100196884302735739144E_4^{15}E_6^6
+387366175916998061434246E^{12}E_6^8
\no\\
&&
+128228189072561942529160E_4^9E_6^{10}
+387366175916998061434246E_4^{12}E_6^8
\no\\
&&
+128228189072561942529160E_4^9E_6^{10}
+304427514534730640824E_4^3E_6^{14}
\no\\
&&
+371716666584611521E_6^{16})
\no\\
&=&q^{-9}+480q^{-8}+61944q^{-7}+1061784q^{-6}+9424164q^{-5}+64348440q^{-4}
+371270060q^{-3}+
\no\\
&&
1864662168q^{-2}+8608773987q^{-1}+134979109134912+\cdots.
\end{eqnarray}
Above results suggest that self-dualized $D,E$ partition functions
does not satisfy the gap condition, except for $D_3$ and $D_5$.
For $D_3$, since $D_3=A_3$ and $A_{N-1}$ series satisfy the gap condition,
this fact is natural. For $D_5$, we have to give some comments.
$D_5$ partition function in (\ref{d5})
seems to satisfy the gap condition.
From the previous discussion, 
the top $q$ term of $D_5$ partition function
must be $q^{-6}$.
Thus this partition function in (\ref{d5}) does not satisfy the gap condition 
in true sense.
Further analysis of $D_{2N+1}$ suggests that
the top $q$ term of the partition function
starts from $q^{-4N}$.
Thus the top $q$ term of $D_{4N+1}$ partition function
may vanish.
This phenomenon
is interesting, but we have no interpretation for this by now.

\subsection{Fulfilling the Gap Condition of Self-Dualized $D,E $ Partition Function}

We assume that holomorphic anomaly does not appear in $D,E $ partition function
because of the following reason.
Holomorphic anomaly comes from the failure of the reduction ${\cal N}=4\to {\cal N}=1$.
Typically in $\frac{1}{2}K3$ case, this failure happens.
By using the mass perturbation proportional to the section of the canonical bundle,
we reduce  ${\cal N}=4\to {\cal N}=1$ \cite{vafa-witten, m-v, E-S}.
But in $\frac{1}{2}K3$ case, there is a locus where
the section of this canonical bundle vanishes.
In this locus, the holomorphic anomaly or $E_2$-depndence comes in
the partition function.
On the other hand, $K3$ has trivial canonical bundle,
that is, there is no locus where the section of this canonical bundle vanishes.
This situation is the same, whether self-dualization is done or not.
We think of the self-dualized $D,E $ partition functions
and intend to satisfy 
the gap condition of these partition functions .
We want to cancel the negative $q$ powers
of the previous partition functions
except for top $q$ term, by using meromorphic modular forms.
In the previous section, we observed that the holomorphic
part of $G_3,G_5$ and $G_7$ are Hecke transformation of $G_1$ plus (polynomial of $j$ function$)\times G_1$.
In the same way as this observation,
we expect the total $D,E$ partition functions has the following form.
The total $D,E$ partition function contains
the self-dualized $D,E$ partition function as a core part
and (polynomials of $j$ function $)\times {\hat Z}_1$ as 
a residual part.
That is,
we assume that top $q$ term only comes from the self-dualized $D,E$ partition function.
These two parts are both holomorphic.
Finally we derive the following results:
\beq
Z_{UD_2}^{tot}(\tau)={\hat Z}_{UD_2}(\tau)+24j{\hat Z}_1(\tau)-30732{\hat Z}_1(\tau)={\hat Z}_3(\tau)=q^{-3}+122976+\cdots,
\enq
\begin{eqnarray}
Z_{UD_4}^{tot}(\tau)&=&{\hat Z}_{UD_4}(\tau)-504j^2{\hat Z}_1(\tau)+1004064j{\hat Z}_1(\tau)-360585162{\hat Z}_1(\tau)
\no
\\
&=&{\hat Z}_5(\tau)=q^{-5}+1575504+\cdots,
\end{eqnarray}
\beq
Z_{UD_5}^{tot}(\tau)=\frac{1}{48}{\hat Z}_{UD_5}(\tau)={\hat Z}_4(\tau)=q^{-4}+532728+\cdots,
\enq
\begin{eqnarray}
Z_{UD_7}^{tot}(\tau)&=&\frac{1}{2}{\hat Z}_{UD_7}(\tau)+1254j^3{\hat Z}_1(\tau)
+3430944j^2{\hat Z}_1(\tau)
-2508200672j{\hat Z}_1(\tau)
+386059650048{\hat Z}_1(\tau)
\no\\
&&
={\hat Z}_8(\tau)=q^{-8}+17047800+\cdots,
\end{eqnarray}
\begin{eqnarray}
Z_{UE_8}^{tot}(\tau)&=&{\hat Z}_{UE_8}(\tau)-480j^7{\hat Z}_1(\tau)
+2679816j^6{\hat Z}_1(\tau)
-5777888472j^5{\hat Z}_1(\tau)
+6037910892540j^4{\hat Z}_1(\tau)
\no\\
&&
-3141074576248824j^3{\hat Z}_1(\tau)
+747778700988800308j^2{\hat Z}_1(\tau)
\no\\
&&
-64381310276022766872j{\hat Z}_1(\tau)
+1077116103457676370261{\hat Z}_1(\tau)
\no \\
&&={\hat Z}_9(\tau)=q^{-9}+29883672+\cdots.
\end{eqnarray}
Resulting gapful partition function are all coincident with
the $U(N)$ partition functions, which 
have the same top $q$ terms.
The characteristics of groups are lost.
This phenomenon is interpreted as universality of gapful partition functions.
If we separate the corresponding $U(1)$ flux part
from the total gapful
$D,E$ partition functions,
we find the difference from the $U(N)$ partition functions.
Unfortunately we have no technique to separate 
$D,E$ partition functions 
from the total gapful
$D,E$ partition functions.

\section{Conclusion and Discussion}
\label{sec:5}
\setcounter{equation}{0}
In this article, we derived the self-dualized $D,E$ partition functions,
which satisfy the gap condition.
The self-dualization was first introduced to construct the holomorphic $U(N)$
partition function
from the $U(1)$ flux part and $SU(N)/{\bf Z}_N$ partition functions, 
by using close relation between 
the orbit structure and
modular forms.
By using the self-dualization,
we self-dualized our $D,E$ partition functions on $K3$
and canceled its negative $q$ powers except for top $q$ term.

Remaining problem is to invent the technique
to separate the orbit structure from general modular forms.
If we can invent this technique,
we can separate 
the true $D,E$ partition functions on $K3$
from gapful self-dualized $D,E$ partition functions.
Ultimate aim is to obtain gapful $D,E$ partition functions
without self-dualization.
We hope to find out the technique to do this and
to find
guiding principle like Seiberg-Witten curve or holomorphic anomaly equation 
in  $U(N)$ on $\frac{1}{2}K3$ case.

In fact we had been writing the another article concerned about ${\cal N}=4~ADE$
partition function \cite{sasaki} parallel to this article.
In \cite{sasaki} we revealed a close relation between Hecke operator and $S$-duality of ${\cal N}=4$ Super Yang-Mills on $K3$.
There are advantages and disadvantages in either article.
We hope that these works are incorporated and we approach to our goal
of the determination of ${\cal N}=4~ADE$ partition function on $K3$. 

{\bf Acknowledgment}\\
We would like to thank M.Jinzenji
for simulating discussions and collaborations.
We also thank Prof.N.Kawamoto for carefully reading our manuscript.

\end{document}